\documentclass[twoside,slac_one]{revtex4}
\usepackage{graphicx}
\usepackage{fancyhdr}
\usepackage{amsmath} 
\usepackage{bm}
\usepackage{amsxtra}
\usepackage{amssymb}
\usepackage{amsthm}
\usepackage{latexsym}
\usepackage{lscape}
\usepackage[stable]{footmisc}

\begin{document}

\title{CMS Computing: Performance and Outlook}

%

\author{Kenneth Bloom (for the CMS Collaboration)}
\affiliation{Department of Physics and Astronomy, University of
  Nebraska-Lincoln, Lincoln, NE, USA}

\begin{abstract}
  After years of development, the CMS distributed computing system is now
  in full operation. The LHC continues to set records for instantaneous
  luminosity, and CMS continues to record data at 300~Hz. Because of the
  intensity of the beams, there are multiple proton-proton interactions per
  beam crossing, leading to larger and larger event sizes and processing
  times. The CMS computing system has responded admirably to these
  challenges. We present the current status of the system, describe the
  recent performance, and discuss the challenges ahead and how we intend to
  meet them.
\end{abstract}

\maketitle

\thispagestyle{fancy}

\section{The problem}
Experiments at the Large Hadron Collider (LHC)~\cite{bib:LHC} will
produce tremendous amounts of data.  With instantaneous luminosities
of $10^{34}$~cm$^{-2}$s$^{-1}$ and a crossing rate of 40~MHz, the
collision rate will be about $10^9$~Hz.  But the rate for new physics
processes, after accounting for branching fractions and the like, is
of order $10^{-5}$~Hz, leading to the need to select events out of a
huge data sample at the level of $10^{-14}$.

The Compact Muon Solenoid (CMS) experiment~\cite{bib:CMS} 
developed a distributed computing model from the very early days of the
experiment. There are a variety of motivating factors for this: a single
data center at CERN would be expensive to build and operate, whereas
smaller data centers at multiple sites are less expensive and can leverage
local resources (both financial and human).  But there are also many
challenges in making a distributed model work.

\begin{figure}[ht]
\centering
\includegraphics[width=80mm]{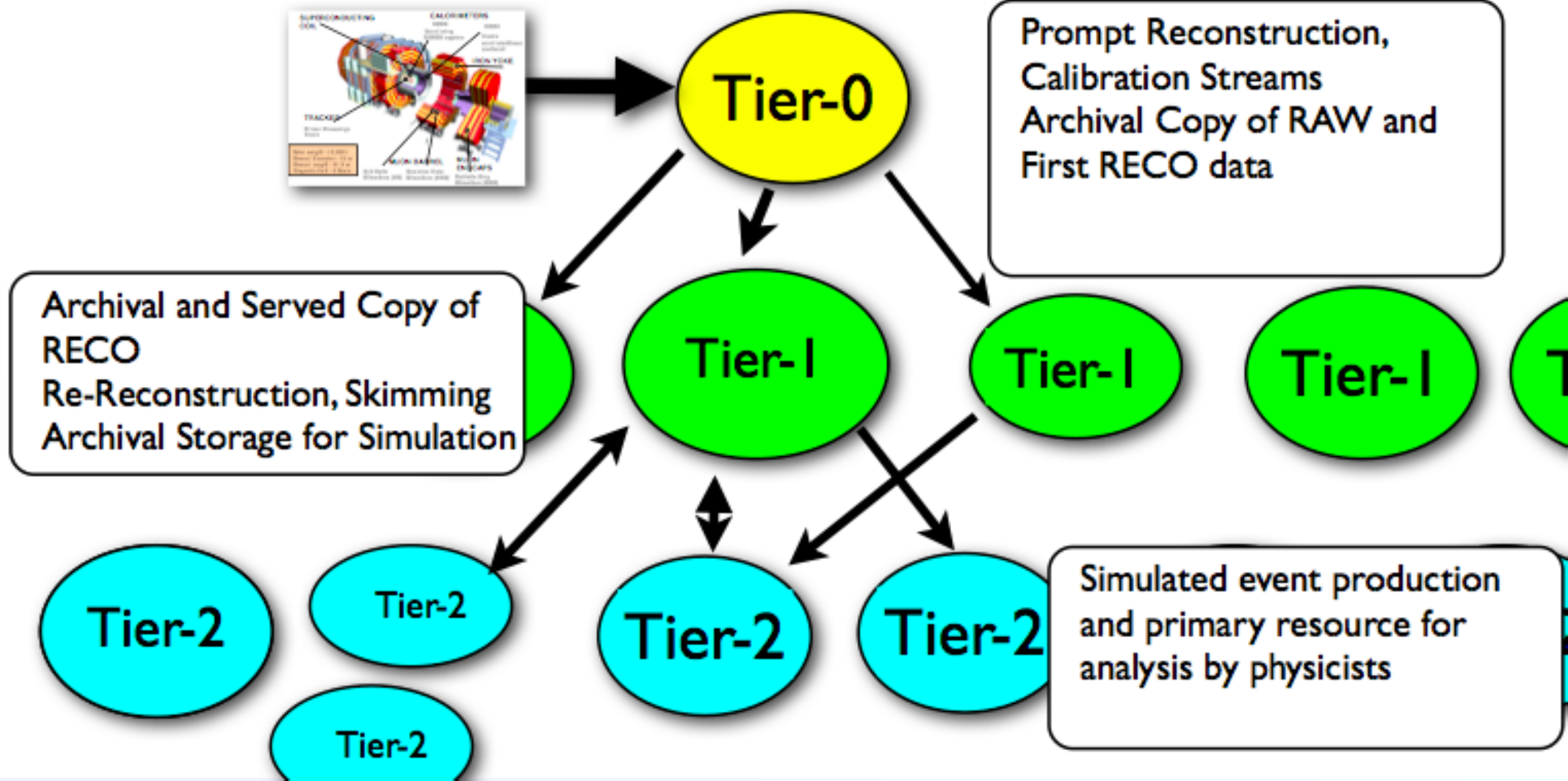}
\caption{Tiered hierarchy of the CMS distributed computing
  model.} \label{fig:tiers}
\end{figure}

The CMS distributed computing model~\cite{bib:CMSmodel} has different
computing centers arranged in a ``tiered'' hierarchy, as illustrated
in Figure~\ref{fig:tiers}, with experimental data typically flowing
from clusters at lower-numbered tiers to those at higher-numbered
tiers.  The different centers are configured to best perform their
individual tasks.  The Tier-0 facility at CERN is where prompt
reconstruction of data coming directly from the detector takes place;
where quick-turnaround calibration and alignment jobs are run; and
where an archival copy of the data is made.  The facility is typically
saturated by just those tasks.  There are seven Tier-1 centers in
seven nations (including at FNAL in the United States).  These centers
keep another archival copy of the data\footnote{If one would say that
  the data is not truly acquired until there are two safe copies of
  it, then the CMS data acquisition system stretches around the
  world.}, and are responsible for performing re-reconstruction of
older data with improved calibration and algorithms, and making skims
of primary datasets that are enriched in particular physics signals.
They also provide archival storage of simulated samples produced at
Tier-2.  There are about 40 Tier-2 sites around the world (including
seven in the U.S.); they are the primary resource for data analysis by
physicists, and also where all simulations done for the benefit of the
whole collaboration take place.  These centers thus host both
organized and chaotic computing activities.

Everything stated so far could just as well have been said two years ago,
before the LHC actually began operations.  In fact, it was~\cite{bib:Bloom}.
What is different now is that we have two years of real operational
experience under our belts.  In this presentation, we discuss the
performance of the computing system in 2010, its anticipated and actual
performance in 2011, the technical advances that have made that level of
performance possible, and some thoughts for the future.

\section{Predicting 2011 from 2010}
2010 was the first full year of LHC operations, and the distributed
computing system performed as expected.  In particular, all workflows ran
at their designated facilities from the very beginning; there was no need
to change Figure~\ref{fig:tiers} on the fly.  The amount of data handled by
the system was truly stunning.  The Tier-0 facility produced 100 different
datasets, with a total of 13.9~billion events and 674~TiB.  At the Tier-1
sites, the data was re-reconstructed 19 times (far more often than
anticipated in the computing model, due to rapidly evolving understanding
of the detector), with 17.2~billion events and 2.4~PiB output.  There were
four re-reconstruction passes on the Monte Carlo (MC) samples, with
8.3~billion events and 2.9~PiB output.  MC production was done at both Tier
1 and Tier 2, with upwards of 500~million events/month produced at peak
rates.

There were also many transfers from tier to tier.  The movement of analysis
datasets to Tier 2 kept up with data-taking, so that data was in the hands
of analysts within a day of being reconstructed.  The original computing
model envisioned peak rates of 600~MB/s from Tier 0 to Tier 1 and 1200~MB/s
from Tier 1 to Tier 2; these were routinely exceeded.  The original model
did not envision transfers among Tier-2 sites, but in fact Tier-2 sites
received about as much data from other Tier-2's as they did from the
Tier-1's.  This created more flexibility and efficiency in how data could
be moved.

Meanwhile, user analysis was successfully migrated to the Tier-2 sites.  It
was impossible to know for sure that the grid could handle hundreds of
users, or even if all of those users would want to use the grid to begin
with.  But in 2010 there were about 450 unique analysis users per week,
submitting 150,000 analysis jobs per day.  The true metric of success was
that, by the time of this conference, 75 papers on the 2010 data had been
submitted, accepted or published, with more in the pipeline, and there was
never any evidence that computing was ever the bottleneck.  It is
remarkable that all of this was achieved in the face of rapidly changing
experimental conditions.

As wonderful as this is, it must be remembered that the LHC only delivered
45~pb$^{-1}$ of integrated luminosity.  This is less than what the
computing system was designed for, and success was mandatory under those
conditions.  However, it provided an opportunity to shake down the system
under a relatively small load, and to gather data that would help plan for
2011.

As the 2011 run approached, it was clear that the target instantaneous
luminosity for the year would be reached rather quickly, and indeed it was
in June.  It was also expected that this would be achieved by having large
proton bunches with many interactions per event; 16 was the amount
anticipated by the September technical stop.  As a result of this, the
event size was expected to double (to 0.8~MB/event for the comprehensive
RECO data format, 0.2~MB/event for the stripped-down AOD format) from the
2010 values, and the processing time was expected to quadruple (to
96~HS06/event)~\cite{bib:HS06}.  While the trigger rate was nominally
expected to be 300~Hz, it was expected that it would be a challenge to keep
it there, given the pressures to try to avoid raising thresholds in the
face of higher event rates.

The experience of 2010 plus the above parameters were used as the basis for
a very thorough modeling effort of the necessary computing resources.  The
expected available resources had already been established through national
pledges to the Worldwide LHC Computing Grid (WLCG)~\cite{bib:WLCG}.  The
model could then be used to tailor operational plans to make sure
activities could fit into the resources available.  What was clear was that
CMS computing was expected to be resource-limited, even after squeezing a
lot of efficiency out of operations.  A few highlights of the modeling
follow.

\begin{figure*}[ht]
\centering
\includegraphics[width=80mm]{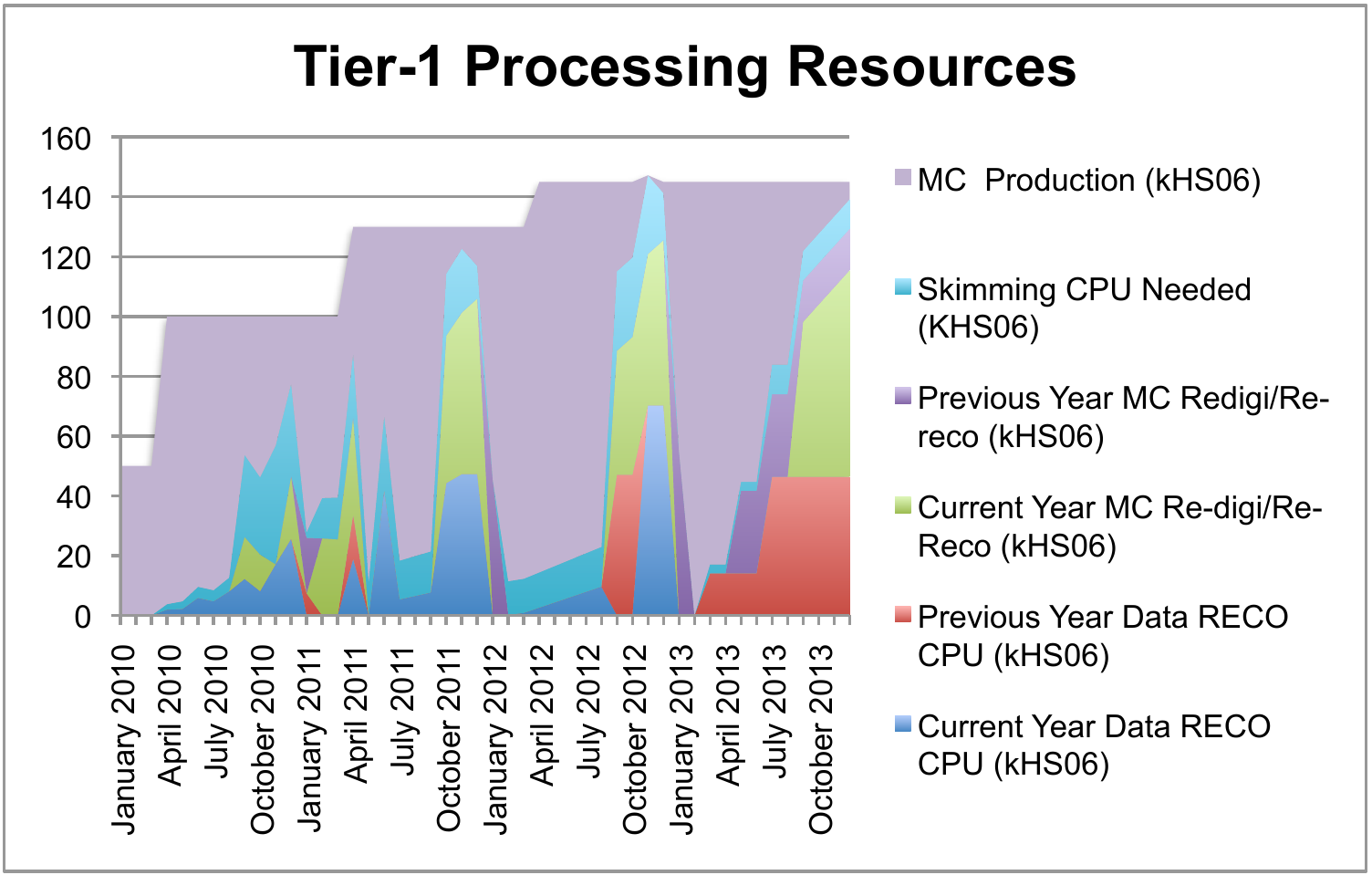}
\includegraphics[width=80mm]{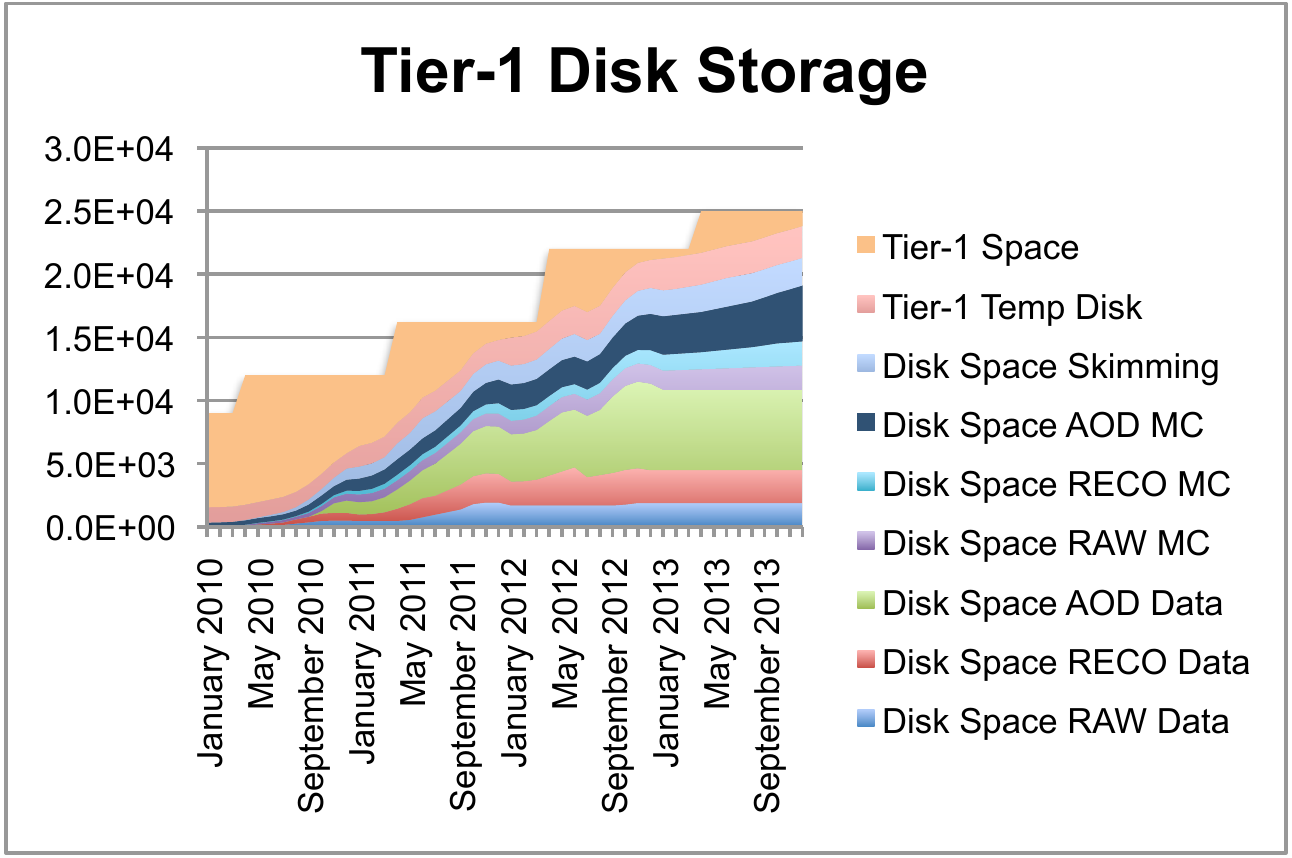}
\caption{Expected use of Tier-1 processing (left) and disk (right) resources by
  month through 2013, broken down by activity.  The step-function envelope
  shows the WLCG pledges for 2011 and CMS requests for 2012 and 2013.} \label{fig:t1capacity}
\end{figure*}

One fact that emerged from 2010 operations (but in retrospect seems fairly
obvious) is that the Tier-1 facilities, built at a scale to handle data
re-processing when necessary, are not very busy when they are not
re-processing data.  Thus, it makes sense to move as much MC production as
possible from Tier 2 to Tier 1, to make use of available Tier-1 resources
and to leave more room for user analysis at Tier 2.  The left panel of
Figure~\ref{fig:t1capacity} shows how different activities are expected to
make use of Tier-1 processing resources month by month through 2013.  One
can see that the Tier-1 centers are extremely busy when re-processing, but
still less so otherwise.  The right panel of Figure~\ref{fig:t1capacity}
shows how disk space is expected to be used at Tier-1.  Less space is
allocated to AOD's at Tier 1 than in the original model, which imagined
that each of the seven centers would keep a complete copy of the AOD.  But
the reliability and speed of data transfers gives confidence that keeping
only two copies of the AOD events across the entire system will be
sufficient.  Still, to fit within the available resources, physicists must
switch from RECO to AOD format as much as possible, and regular deletion
campaigns will be required.

\begin{figure*}[ht]
\centering
\includegraphics[width=80mm]{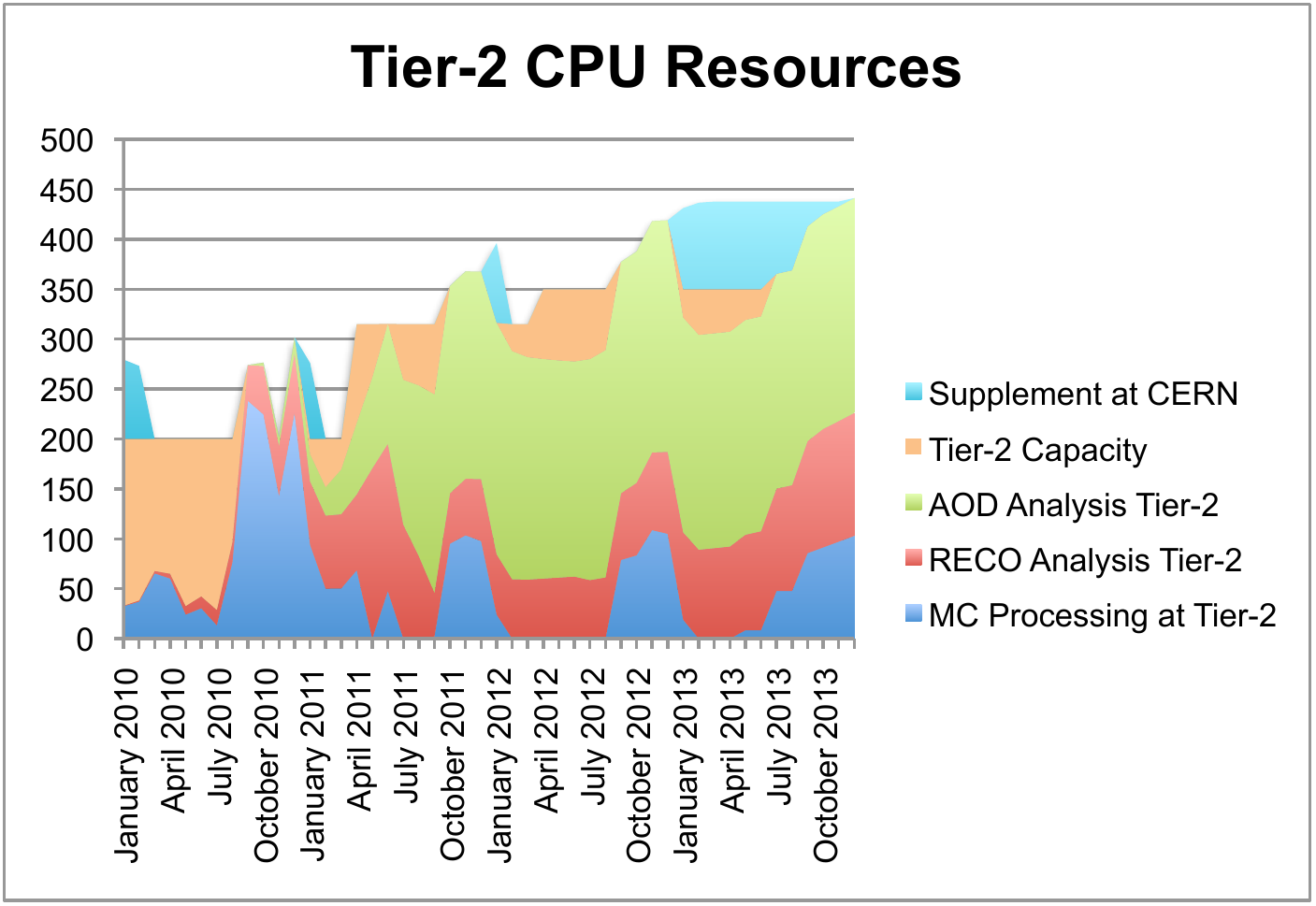}
\includegraphics[width=80mm]{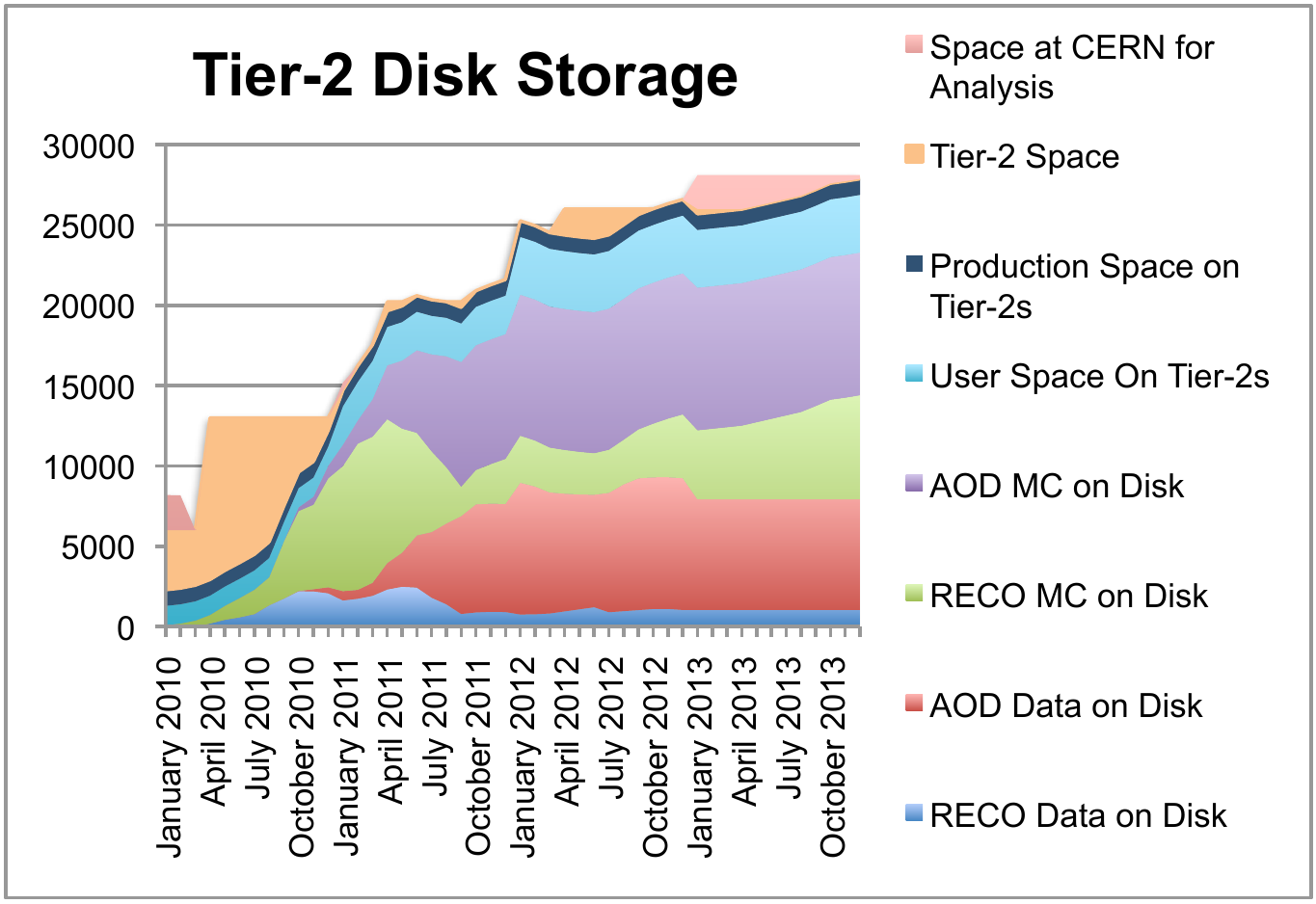}
\caption{Expected use of Tier-2 processing (left) and disk (right) resources by
  month through 2013, broken down by activity.  The step-function envelope
  shows the WLCG pledges for 2011 and CMS requests for 2012 and 2013.} \label{fig:t2capacity}
\end{figure*}

Figure~\ref{fig:t2capacity} shows similar plots for the planned use of
Tier-2 resources.  As long as much MC production activity stays at Tier 1,
the use of processing resources at Tier 2 is reasonable.  But during
re-processing periods, MC production moves back to Tier 2, at which point
the resources are overcommitted.  In addition, to remain within the
available disk resources, 90\% of user analysis needs to move from RECO to
AOD samples.  The model assumes that there will be four copies of each
analysis dataset across all of the Tier-2 sites, but this might have to be
reduced if there is a disk-space crunch.  Under any circumstances, Tier-2
resources are heavily committed over the next few years.

\section{2011 operational experience}

How well does real CMS life match up with the plan, which was developed
before the LHC started running this year?  Here we discuss recent
operational experience, and some of the technological changes that have
been implemented to make the CMS computing system work better.

\subsection{Data size and rate}

\begin{figure}[ht]
\centering
\includegraphics[width=80mm]{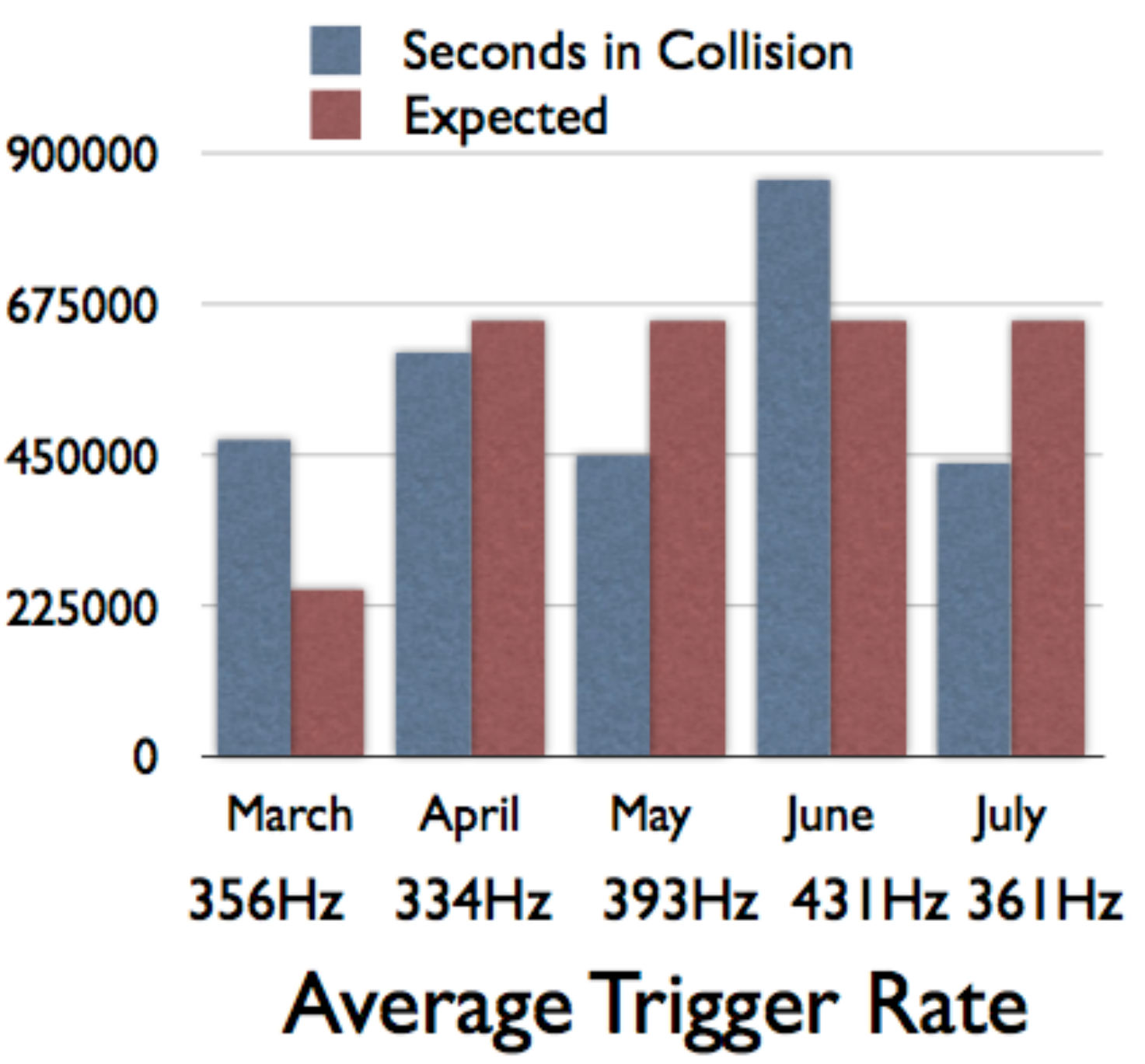}
\caption{Time per month that the LHC provided collision data to
  experiments, compared to the values used in the planning.  The average
  trigger rate in each month is also shown.} \label{fig:runtime}
\end{figure}

Figure~\ref{fig:runtime} shows the amount of time that the LHC has been
operating for physics data-taking each month, through July, compared to the
expected operational time that was an input to the model.  The figure also
shows the average trigger rate.  Overall the LHC duty cycle has been lower
than expected, but this has been compensated for by a trigger rate that is
consistently greater than 300~Hz.  (The trigger rate includes the overlap
in primary datasets, which was planned to be about 25\%.)  In total, about
1.1~B events have been recorded, compared to 1.3~B expected from the
model.  A small amount of contingency has been gained as a result.  Given
this rate, the size of the full 2011 dataset after a re-reconstruction pass
should be about 1~PB.

Table~\ref{tab:eventsize} compares the sizes of different kinds of events
to the expected values, which were based on simulations with realistic
running conditions.  In general, pileup has been lower than anticipated,
due to an optimization of the luminosity that led to adding more proton
bunches rather than increasing the number of protons per bunch.  Event
sizes are smaller than expected as a result.  The time required to
reconstruct events has been about as expected for minimum-bias events, and
about 20\% larger than planned for other datasets.  So far, event sizes and
processing times have been roughly constant with increasing luminosity.
But the LHC has now reached the limit of how many bunches can be circulated
in the current configuration, and luminosity must be increased in ways that
increases the pileup at the same time.  Thus the events are expected to get
larger and processing times longer as the year continues.

\begin{table}[ht]
\begin{center}
\caption{Expected and observed event sizes, in kilobytes.}
\begin{tabular}{cccccc}
Data Type & Data RAW & Data RECO & Data AOD & MC RECO & MC AOD\\\hline
Expectation & 390 & 530 & 200 & 600 & 265\\
Observed & 200 & 500 & 100 & 970 & 250\\\hline
\end{tabular}
\label{tab:eventsize}
\end{center}
\end{table}

\subsection{Tier 0}

\begin{figure*}[ht]
\centering
\includegraphics[width=80mm]{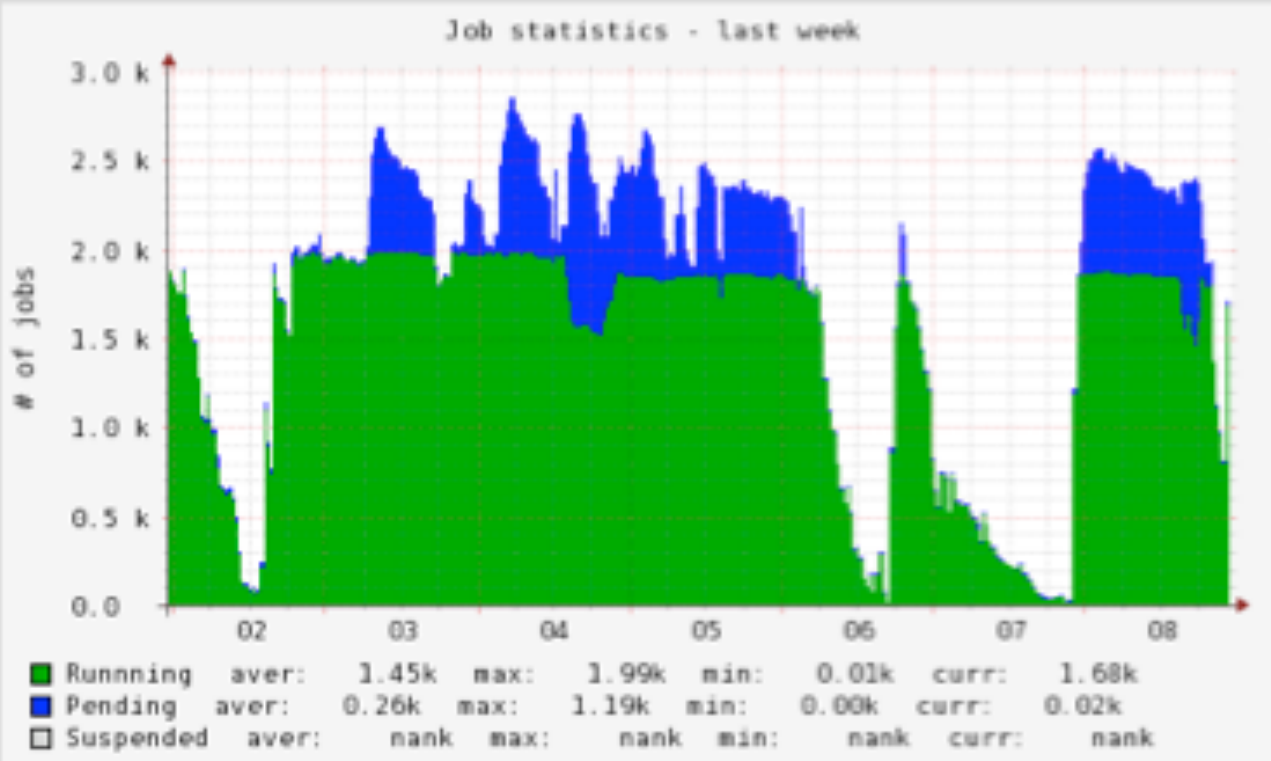}
\includegraphics[width=80mm]{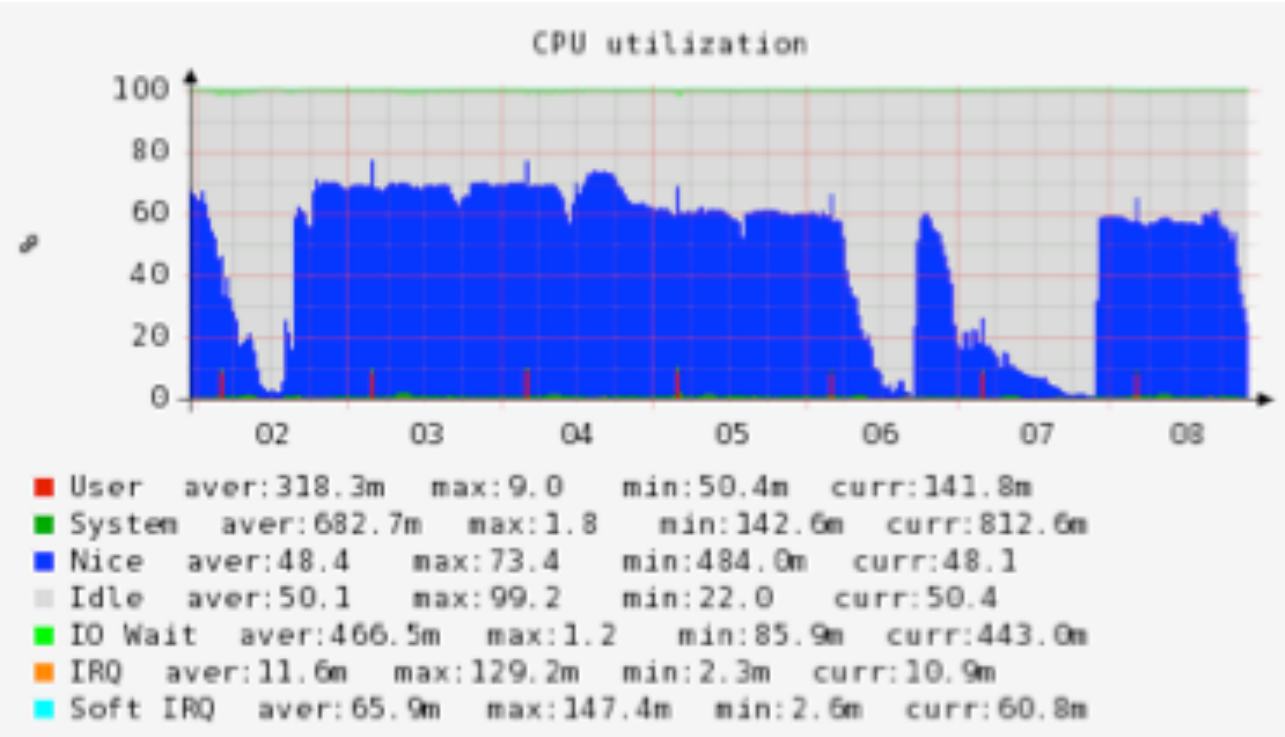}
\caption{Number of CMS jobs running and queued (left) and CPU utilization
  fraction (right) at the Tier-0 facility during June 2011.  The day of the
month is indicated on the horizontal axis.} \label{fig:t0use}
\end{figure*}

As for the operations of facilities at the various tiers, the left panel of
Figure~\ref{fig:t0use} shows the numbers of jobs running and queued during
the first weekend in June, when the LHC had 40\% livetime.  As can be seen,
the Tier-0 cluster was saturated, leading to nearly a thousand jobs queued
at times.  Once the machine stopped, this backlog was quickly cleared.
However, the right panel of the figure shows that the processors were not
fully used.  This is because the switch to 64-bit executables and a new
version of ROOT led to a larger memory footprint that inhibited efficient
use of CPU.  Work is in progress to reduce the size of the executable, and
to take advantage of whole-node scheduling, in which multiple similar jobs
could run on a single node and share read-only memory.  Despite these
challenges, the Tier-0 facility is keeping up well enough with
reconstructing the data as they arrive.

\subsection{Tier 1}

\begin{figure*}[ht]
\centering
\includegraphics[width=80mm]{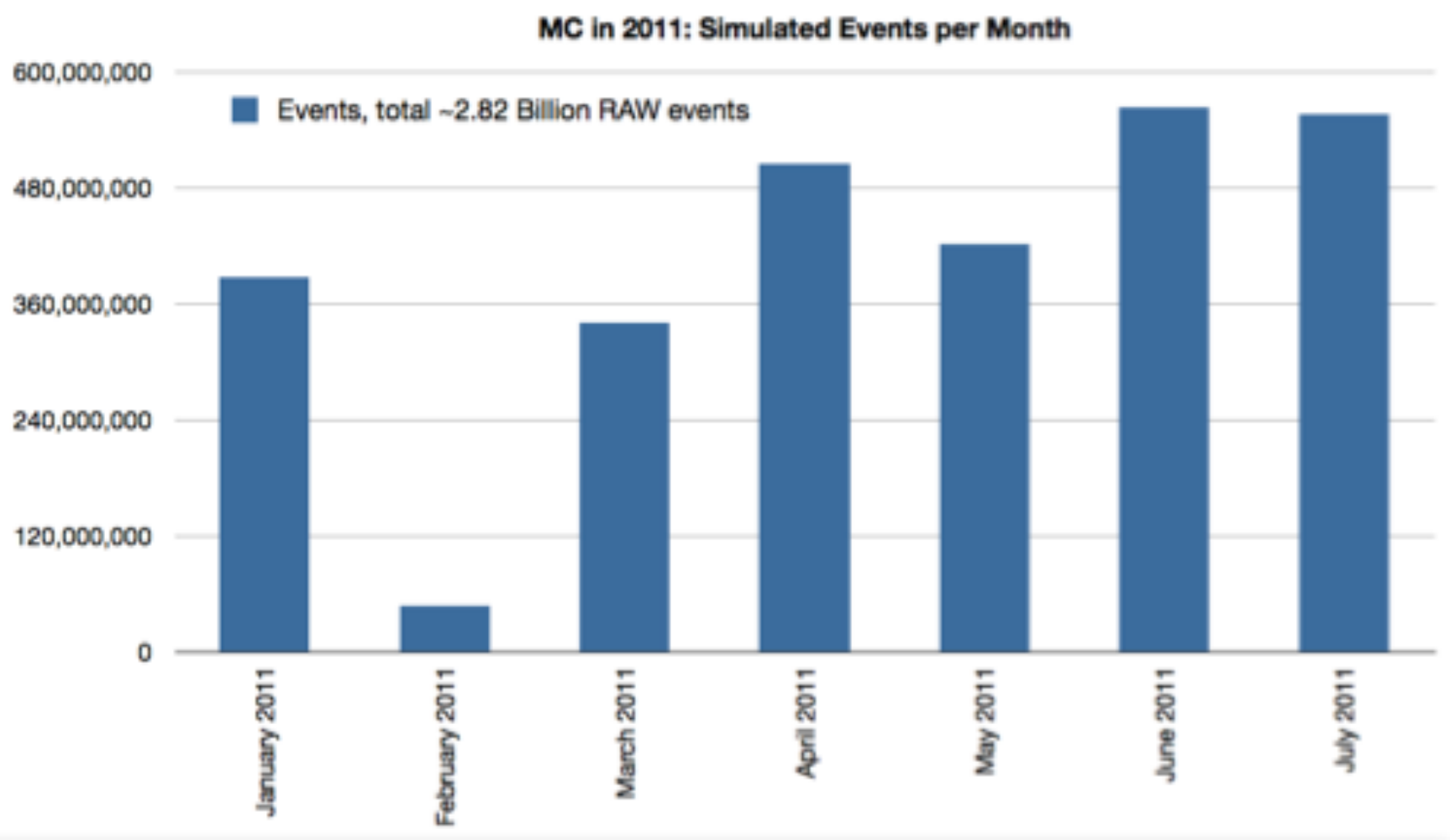}
\caption{Number of Monte Carlo events produced per month in 2011.} \label{fig:mcprod}
\end{figure*}

The seven Tier-1 facilities completed a full re-reconstruction of the 2010
data (nearly 1.5~B events) in April, and all available 2011 data (about
600~M events) in May, as was expected in the original planning.  It is
possible that there will not be another re-reconstruction of the data
before the end of 2011, but this depends on whether the software version at
Tier 0 is changed because of challenging event environments.  Meanwhile,
2.8~B MC events were produced through the end of July, as indicated in
Figure~\ref{fig:mcprod}.  The latest simulation samples include out-of-time
pileup.  The expected production capacity was 0.22~B events/month, but in
fact the system has been capable of much more than that.

Some of the success of the Tier-1 operations can be attributed to new
technology that was implemented this year.  The workflow management system
that had been used for data re-processing was originally designed for MC
production.  In that use case, it is not fatal to lose some of the events
in the course of the processing, as more can always be made.  But it is
unacceptable to lose any data events.

A new workflow management system, WMAgent~\cite{bib:WMAgent}, is much more
robust against such problems, and its deployment was a great help to
operations.  It is a state machine rather than a messaging system, and has
100\% accountability for all events processed.  One issue that arose was
that the current version of the reconstruction software uses more memory
than before, and as a result jobs were running longer and more jobs failed.
Because WMAgent can redo failed jobs straightforwardly, this did not become
an operational hurdle.  Obviously, the system has allowed for more
efficient MC production too.

Work has also begun to implement whole-node scheduling at the Tier-1 sites
for further operational efficiencies.  The goal is to have 50\% of Tier-1
resources used this way by the end of the year.

\subsection{Tier 2}

\begin{figure*}[ht]
\centering
\includegraphics[width=50mm]{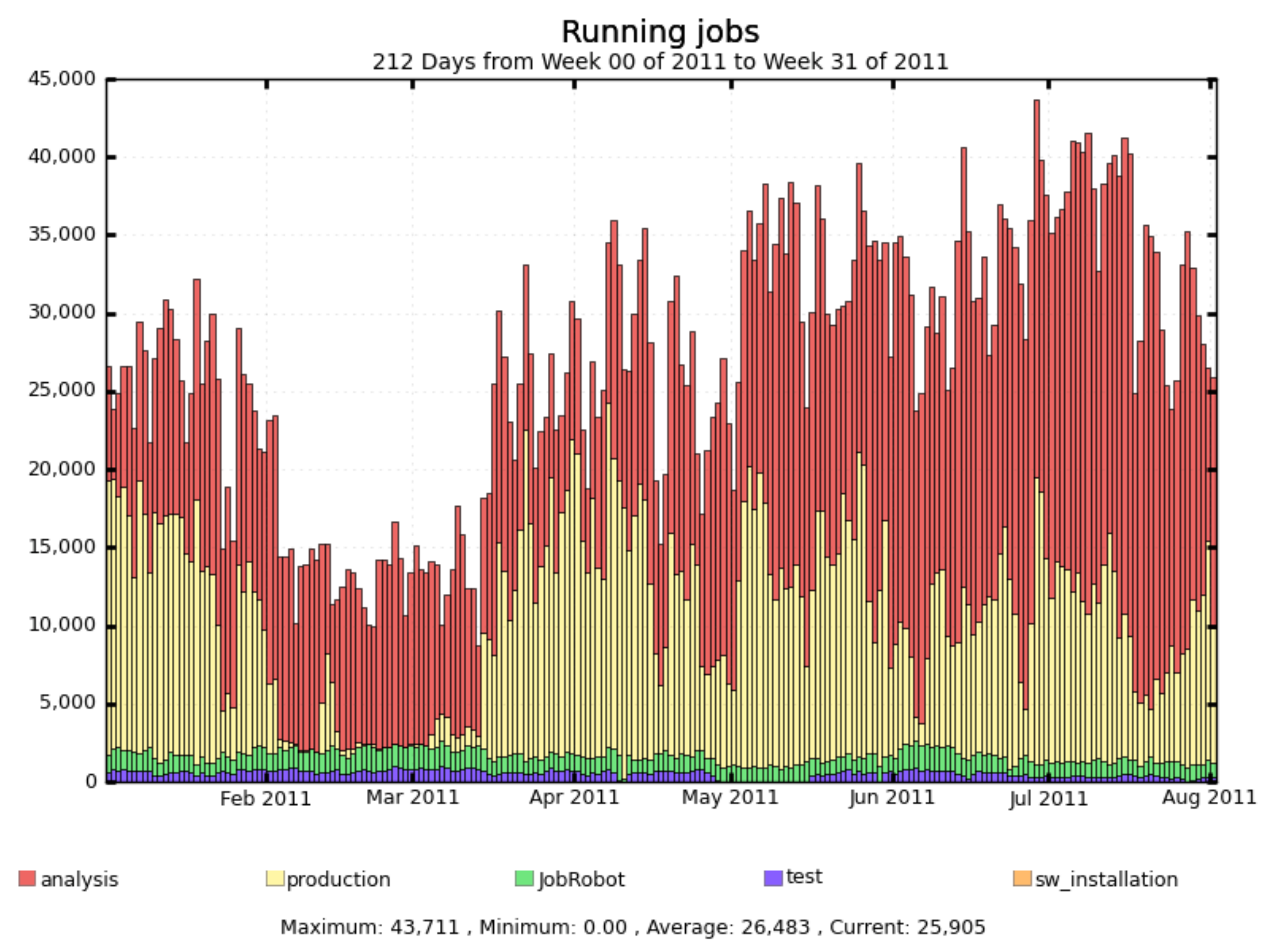}
\includegraphics[width=50mm]{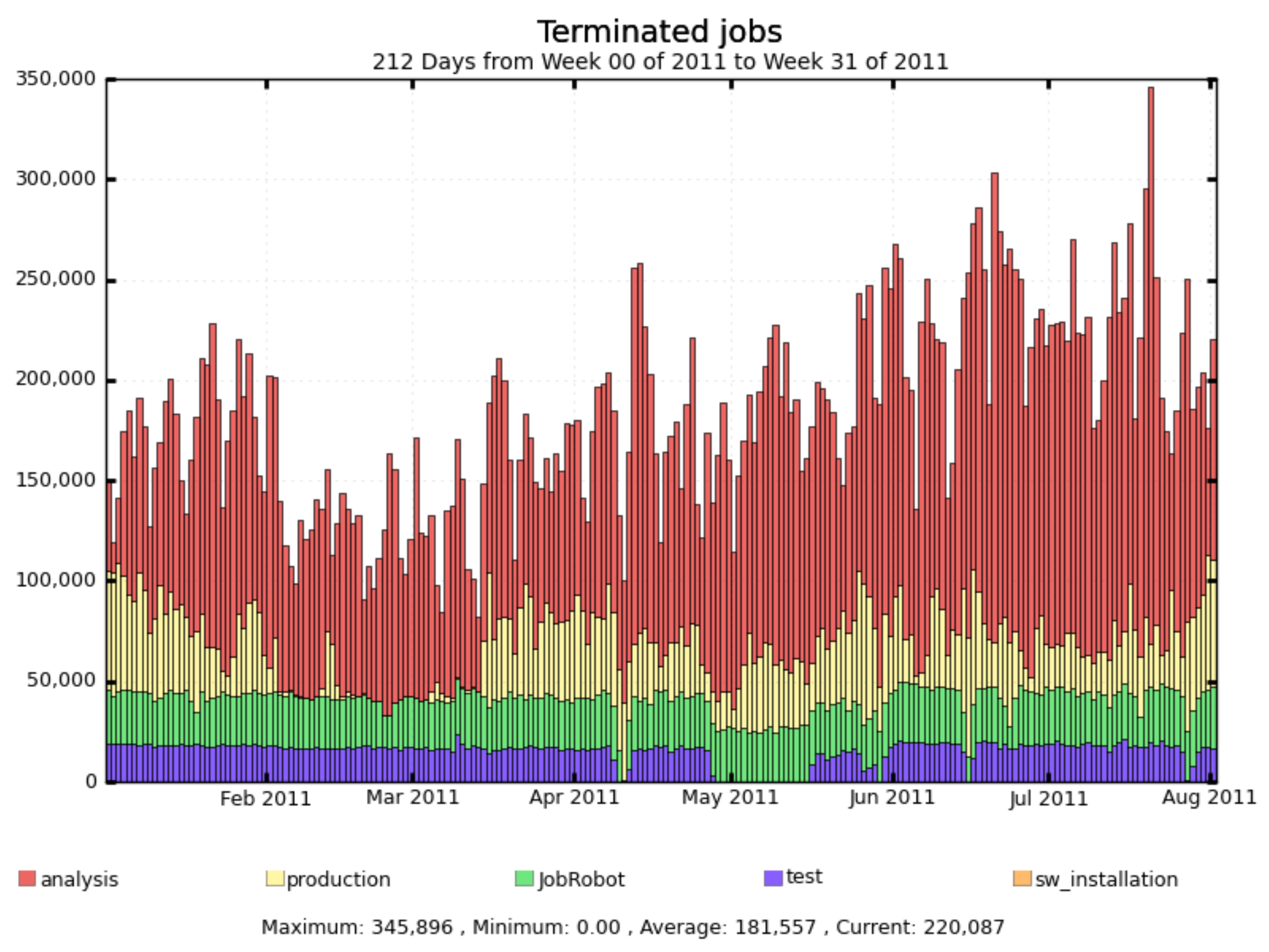}
\includegraphics[width=50mm]{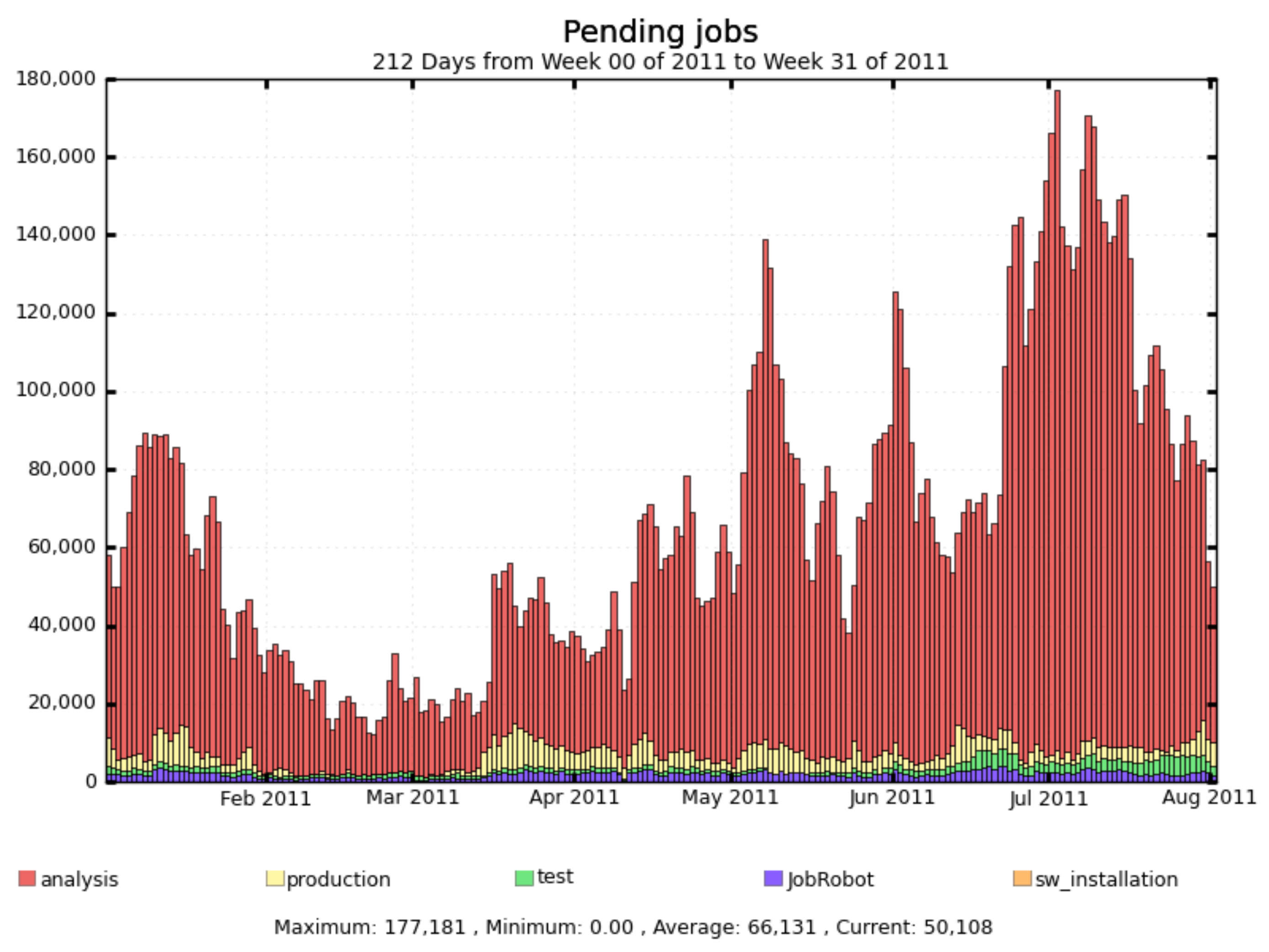}
\caption{Number of jobs running (left), completed (center) and pending
  (right) at CMS T2 sites through the first seven months of 2011.} \label{fig:t2use}
\end{figure*}

Figure~\ref{fig:t2use} shows the number of running, completed and pending
jobs at the approximately 50 CMS Tier-2 centers during the year so far.
About 30,000 cores are continually available for use, with an increasing
number of them devoted to analysis use, as more MC production moves to Tier
1.  About 250,000 grid jobs are completed each day, more than was
anticipated in the original computing model.  Still, thousands of jobs are
pending on any given day, with longer queues at times of great analysis
activity.  This suggests that more resources are needed for analysis,
and/or that the jobs are not being optimally scheduled across all the
sites.

\begin{figure*}[ht]
\centering
\includegraphics[width=80mm]{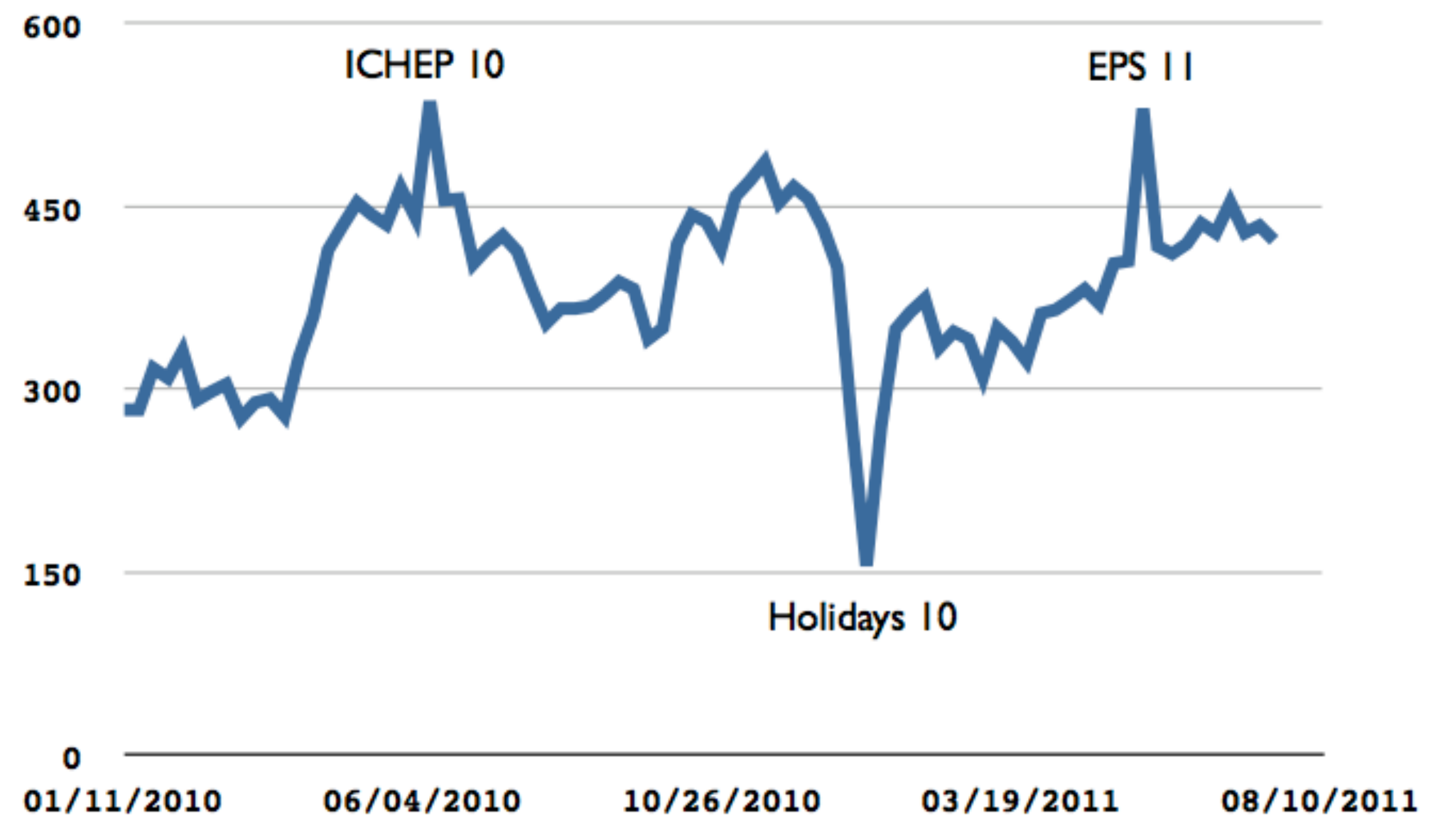}
\caption{Number of unique CMS grid users by week since the start of 2010.} \label{fig:anausers}
\end{figure*}

Meanwhile, the user community continues to grow.  Figure~\ref{fig:anausers}
shows the number of unique analysis users in the CMS distributed computing
system over the past year and a half.  It is steadily growing, modulo peaks
and valleys that can be correlated with important events on the
particle-physics calendar.  A significant fraction of the collaboration,
about 800 users/month, is making use of grid resources.

\begin{figure*}[ht]
\centering
\includegraphics[width=80mm]{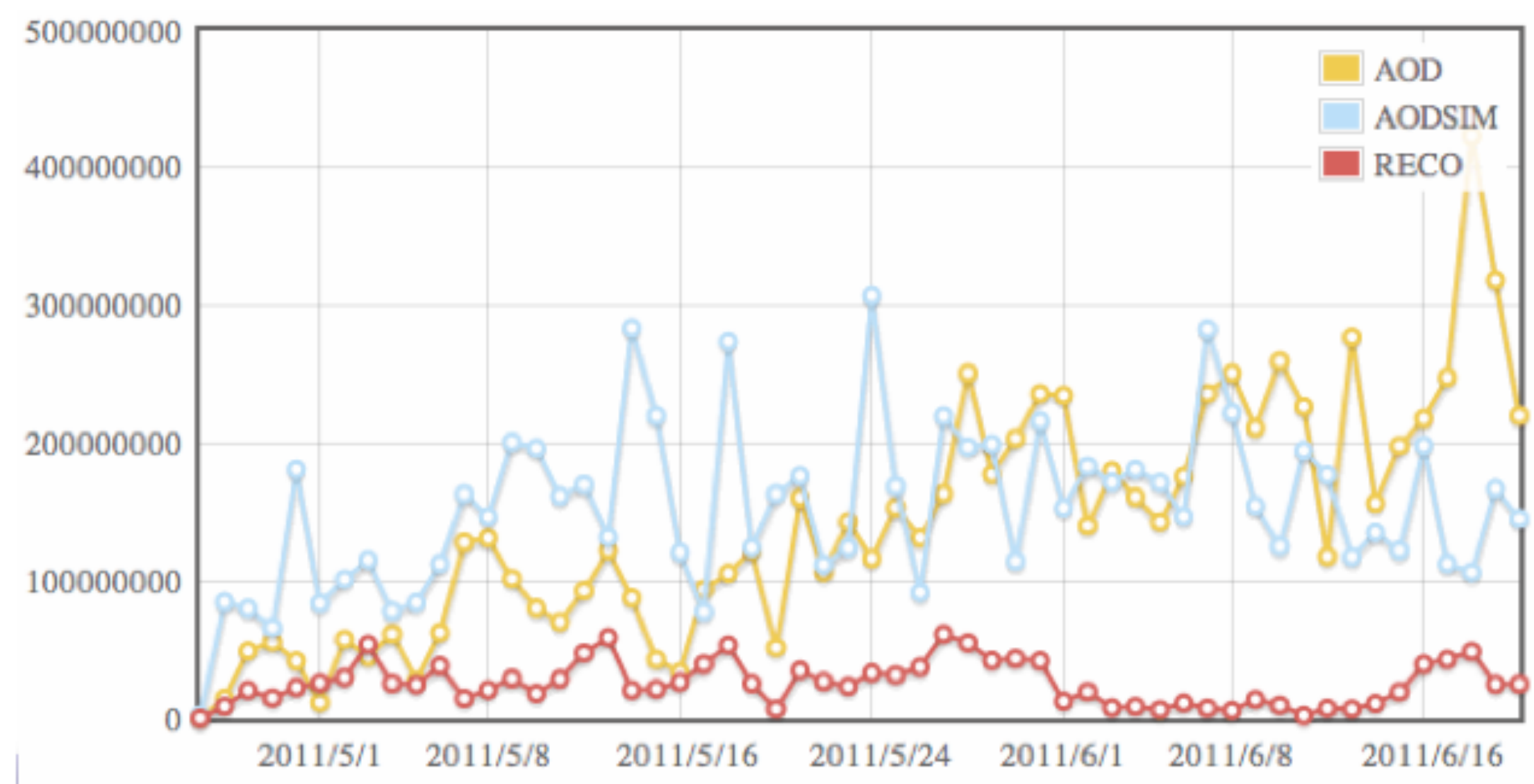}
\caption{Evolution in use of analysis datasets during 2011.  The increased
  use of AOD and AODSIM and declining use of RECO is important for CMS to
  remain within its available computing resources.} \label{fig:dataset}
\end{figure*}

As noted in the previous section, it was imperative that analysis users
start to move towards using the more compact AOD format for their work CMS
was to stay within the available computing resources.  Tools have recently
been developed to track dataset usage in greater detail.  These tools
indicate that the migration is indeed happening as planned, as shown in
Figure~\ref{fig:dataset}.  Another experiment previously had these tracking
tools; CMS is looking forward to using them to help manage dataset
distribution and more.

The operation of the grid sites for analysis will improve with some
technology developments that will be deployed shortly in CMS.  The CMS
Remote Analysis Builder (CRAB)~\cite{bib:CRAB} that analysts use to submit
jobs will soon have a significant revision.  WMAgent will be installed
underneath to take advantage of its features.  As a result, the user
interface will change, requiring some user re-education.  Also, the greater
use of pilot jobs (or ``glide-ins'') for analysis is anticipated.  This
could allow for a prioritization of user jobs across the distributed
system, not just at individual sites, and could also have potential for
balancing usage across sites.

\section{Any Data, Anytime, Anywhere\footnote{The author of this paper
    coined this phrase, and continues efforts to firmly establish the primacy of his claim.}}
One fact that has emerged from the operation of the CMS distributed
computing infrastructure for data analysis is that a key limitation of the
computing model is that CPU and storage must be co-located; a dataset must
effectivly reside on a disk that is in the same room as the compute node
that runs the program that analyzes the dataset.  Thus, the data must be
placed where the processing resources are.  This is difficult to optimize,
as that relies on having some sense of analyst preferences of datasets and
sites.

However, we now know that wide-area networking is more reliable than was
anticipated when the original MONARC model~\cite{bib:MONARC} was developed.  There, dataset
transfer between sites was avoided as much as possible.  At the same time,
CMS has made much progress in optimizing the reading of data files over the
network, so that there is very little additional cost compared to reading a
file in the same room.

Thus, one is inclined to forget co-location and to think big.  What if
users could analyze data in one place with a CPU that is in another place?
In such a scheme, data placement would hardly matter anymore, as any data
would be available anytime, anywhere.  Users could be insulated from
storage problems at sites; if a file was corrupt at one site, there could
be a straighforward and quiet failover to the newtork for access of the
same file at a different site.  Participation in data analysis could be
broadened by enabling users who do not have large storage systems, as those
users could still have access to any data.  The dream of a ``diskless
Tier-3'' site becomes realistic.  Also, it would be straightforward to
access data with cloud resources, should that become cost-competitive.

Prototype systems for such a scheme have already been deployed, using the
Xrootd technology~\cite{bib:xrootd}.  A key element is redirectors that
allow jobs to find data at remote sites without any action from the user.
The US CMS Tier-2 sites have been configured so that a failed file access
at a site will fall back to reading the file from another site using the
redirector over the wide-area network.  There is still much work to be done
to test and operate the system at the needed scale, and to develop
monitoring, accounting and throttling systems.  In related work, CMS is
also exploring how to migrate jobs between sites to optimize the use of
processing resources.

\section{Remarks and outlook}

The actual use of CMS computing in 2011 has been largely in line with the
model that was created based on 2010 experience.  Some of the parameters
have ended up being higher or lower, within about 20\%, but the variances
have tended to compensate each other.  The model does predict that CMS will
be limited by its computing resources during this year.  Early indications
of these limitations were 
already being seen in the run-up to this summer's conferences.  Some
analyses were slowed by the wait for simulation samples, and there has been
significant demand for processing resources at Tier 2.  If CERN chooses to
run the LHC at very high luminosity, this could get worse still.  At this
writing, in mid-September 2011, the situation is still quite fluid.
Physicists will need to adapt to this new environment.  However, all of
this should be taken as good news -- the resource limitations reflect the
fact that the LHC datasets are growing rapidly, and provide the opportunity
for the discovery of new physics.

We can conclude that 2010 was an extremely good year for CMS computing.
The distributed system was a strategic asset for producing physics results,
and no one has ever complained that their work was limited by the available
computing.  It is important to keep in perspective that the operational
scales of everyday operations were considered ``bleeding edge'' just a few
years ago.  This strong performance has continued in 2011, but CMS has now
entered an era of resource constraints.  Fortunately, continuing technology
developments have given some operational breathing room, and some of these
advances have the potential to change the paradigm of computing at the LHC,
and for data-intensive, high-throughput computing in general.

\begin{acknowledgments}
  I thank Ian Fisk for his advice in preparing the talk and his general
  comprehensive knowledge of CMS computing.  I also thank the organizers of
  the DPF 2011 conference for an engaging and enjoyable week.
\end{acknowledgments}

\bigskip 

\end{document}